# The application of encoder-decoder neural networks in high accuracy and efficiency slit-scan emittance measurements

S. Ma,[1,2,†] A. Arnold,[1] P. Michel,[1] P. Murcek,[1] A. Ryzhov,[1] J. Schaber,[1] R. Steinbrück,[1] P. Evtushenko,[1] J. Teichert,[1,*] W. Hillert,[2,3] R. Xiang,[1] and J. Zhu[3]

1. Helmholtz-Zentrum Dresden-Rossendorf, 01328 Dresden, Germany

2. Universität Hamburg, 20148 Hamburg, Germany

3. Deutsches Elektronen-Synchrotron (DESY), 22607 Hamburg, Germany.

**Abstract**

A superconducting radio-frequency (SRF) photo injector is in operation at the electron linac for beams with high brilliance and low emittance (ELBE) radiation center and generates continuous wave (CW) electron beams with high average current and high brightness for user operation since 2018. The speed of emittance measurement at the SRF gun beamline can be increased by improving the slit-scan system, thus the measurement time for one phase space mapping can be shortened from about 15 minutes to 90 seconds. A parallel algorithm and machine learning have been used to reduce the beamlet image noise. In order to estimate the uncertainty in the calculation of normalized emittance, we analyze the main error contributions such as slit position uncertainty, image noise, space charge effects and energy measurement inaccuracy.

1. **Introduction**

High gradient radio-frequency photoelectron injectors (rf guns) are a kind of best candidates to generate electron beams with low emittance, high brightness, and ultrashort bunches of pico- to sub-picoseconds at present [1]. The remarkable high beam quality of these electron injectors has opened up new development of modern accelerator-based scientific facilities, such as extreme ultraviolet (XUV) [2,3] and X-ray free-electron lasers (X-FEL) [4], energy recovery linacs (ERL) [5], MeV ultrafast electron diffraction (UED) [6], and hadron-beam cooling [7].

One of the most critical parameters characterizing rf guns is the transverse emittance. Originally, the transverse emittance is defined as the projected area occupied by the beam particles on the two-dimensional trace space xx' [8, 9] at a given position z along the beam axis. The area calculation is difficult for particle distributions with vaguely defined boundaries, and it requires further specifications. An alternative and widely used way to overcome this difficulty is to adopt the concept of root-mean-square (RMS) emittance based on the second moments of the particle distribution, proposed by Lapostolle [10] and Sacherer [11].

† Corresponding author: s.ma@hzdr.de;
* j.teichert@hzdr.de.

How to measure the transverse emittance with high accuracy and speed is always an important question, especially at a user facility where the time slot for the beam diagnostics is limited, and the task should be routinely done by the shift staff. Several measurement methods are popular at most facilities, for instance, multi-monitor measurement [12], quadrupole or solenoid scan [13], beam tomography [14], slit-scan and pepper-pot [15]. An essential factor for selecting a suitable method is that the beam is emittance or space-charge dominated. A criterion is defined in ref. [16]. For the low energy, high brightness beams from photo injectors, the space charge contributes so considerably that such methods like slit-scan or pepper-pot methods should be applied.

At the ELBE radiation center [17], a SRF gun has been developed and operated as an injector for the ELBE electron accelerator since 2007 [18]. At present, the SRF gun of second generation is working, referred as SRF gun-II. It includes a 3.5-cell 1.3-GHz superconducting niobium cavity, a superconducting solenoid, and a photocathode with a corresponding laser system. A cathode supporting system cooled with liquid nitrogen ($LN_2$) allows the operation of normal-conducting photocathodes with a high quantum efficiency (QE). The SRF gun operates in CW mode with a repetition rate up to 13 MHz. A diagnostics beamline allows the detailed characterization of the electron beam and the further optimization of the SRF gun's operation.

To meet the requirement of emittance measurement in our user facility, i.e. speediness and accuracy, we upgraded the slit-scan setup in the diagnostic beamline. The ultra-high-vacuum (UHV) translation stage for the slit movement, drive motors, and control units have been replaced by advanced systems. The slit moving velocity with the new motor is now adjustable from 0 to 25 mm/s. New control software has been developed, and the data analysis methods have been improved meanwhile. Considering that machine learning (ML) has become one of the most widely used and successfully developed method in image processing [19], the benefits expected here for the data analysis of the slit-scan measurement are an increase in speed, a higher accuracy of the beamlet profile, and the omission of manual intervention during the image data processing. Since the noise reduction in the beamlet images is independent, a parallel algorithm can be used to compress the processing time further.

In the first part of this paper, the theory of the slit-scan emittance measurement method is briefly reviewed. The second part describes the emittance measurement system in the SRF gun beamline. The third part introduces two convolutional neural networks (CNN) [19-21], whereby one is for beamlet images classification, and the other one, named encoder-decoder CNN, is for image noise reduction. The measurement error of the slit-scan process is discussed and analysed in the fourth part. Computer simulation of the slit-scan method based on the ASTRA code [22] is applied for the analysis in this part. The following fifth part presents our emittance measurement results with the new slit-scan system. The preciseness of the new measurement system is confirmed by comparing the experimental results with ASTRA beam dynamics simulation. The conclusions are given in the last part.

## 2. Slit-scan technique

### 2.1. Emittance measurement

The slit-scan and multi-slit techniques are widely used for measuring the transverse phase space and transverse projected emittance of high brightness electron beams produced by photo injectors. In these methods, the space charge dominated beam is split into many small and emittance dominated beamlets using a mask with one or several narrow slits. The particles of the beamlets drift from the mask position to a screen, as shown in Fig. 1. The data analysis works in the same way for single-slit scan as for multi-slit scan method. In both cases the particle divergence is transferred to a position distribution.

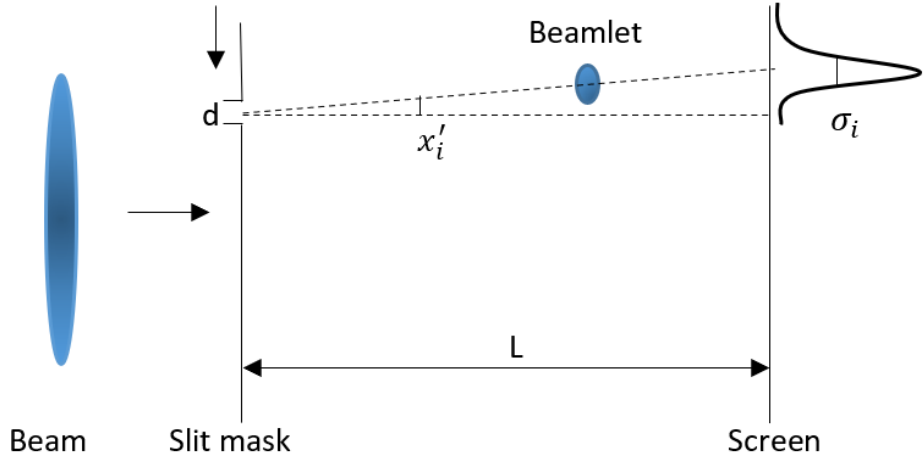

FIG. 1. Illustration of single-slit scan emittance measurement scheme.

For the beamlets a linear beam transport between the slit mask is assumed and thus the observation screen can be described by the transport matrix

$$\begin{pmatrix} x_{screen} \\ x'_{screen} \end{pmatrix} = \begin{pmatrix} 1 & L \\ 0 & 1 \end{pmatrix} \begin{pmatrix} x_{slit} \\ x'_{slit} \end{pmatrix}. \tag{1}$$

The RMS transverse emittance $\varepsilon$ is defined as

$$\varepsilon = \sqrt{\langle x^2\rangle\langle x'^2\rangle - \langle xx'\rangle^2} , \qquad (2)$$

and a normalized transverse emittance $\varepsilon_n$ can be written as

$$\varepsilon_n = \frac{p_z}{m_0 c}\sqrt{\langle x^2\rangle\langle x'^2\rangle - \langle xx'\rangle^2}. \qquad (3)$$

Here the second moments $\langle x^2\rangle$, $\langle x'^2\rangle$, $\langle xx'\rangle$ are given based on the particle distribution $\rho(x, x')$:

$$\langle x^2\rangle = \int \rho(x-\langle x\rangle)^2 dx dx', \langle x'^2\rangle = \int \rho(x'-\langle x'\rangle)^2 dx dx',$$

$$\langle xx'\rangle = \int \rho(x-\langle x\rangle)(x'-\langle x'\rangle)dx dx', x' = \frac{p_x}{p_z} , \qquad (4)$$

where $p_x$ is the horizontal beam momentum, $p_z$ is the longitudinal beam momentum and it is approximately equal to the beam momentum, $m_0$ is the rest mass of an electron, and $c$ is the speed of light. In the slit-scan case, the $\langle\ \rangle$ is related to an averaging over the beamlets and $n_i$ is the particle intensity through the slit at $i$-th position. The bunch center at slit position is $\langle x\rangle = \frac{\sum n_i x_{si}}{\sum n_i}$, and $x_{si}$ is the slit coordinate at $i$-th position. The beam size at the slit position is $\langle x^2\rangle = \frac{\sum n_i(x_{si}-\langle x\rangle)^2}{\sum n_i}$. The bunch average divergence is $\langle x'\rangle = \frac{\sum n_i \overline{x'}_i}{\sum n_i}$, and $\overline{x'}_i$ is the $i$-th slit position's average divergence which can be obtained by $\overline{x'}_i = \frac{\langle \overline{x}_{sci}\rangle - x_{si}}{L}$. Here $\langle \overline{x}_{sci}\rangle$ is the position of the $i$-th beamlet center on the screen, and $L$ is the drift distance. The $i$-th beamlet divergence is $\sigma_i'^2 = \frac{\sigma_i^2 - \frac{d^2}{12}}{L^2}$, $\sigma_i$ is the $i$-th beamlet RMS size on the screen, $d$ is the slit size and the factor 12 is from the RMS value of the slit. The quantities in Eq. (3) can be expressed by the values measured at the screen:

$$\langle x'^2\rangle = \frac{\sum[n_i \sigma_i'^2 + n_i(\overline{x'}_i - \langle x'\rangle)^2]}{\sum n_i} , \qquad (5)$$

$$\langle xx'\rangle = \frac{\sum n_i x_{si}\overline{x'}_i}{\sum n_i} . \qquad (6)$$

### 2.2. Fast slit-scan measurement system at ELBE

The slit-scan system is a part of the diagnostics beamline, as shown in Fig. 2 [18]. The whole system is located at screen-station 2 and 3. Station 2 includes one single-slit mask with a 100 µm wide slit, one yttrium aluminum garnet (YAG) screen, and one calibration screen. Station 3 has one YAG screen, one optical transition radiation (OTR) screen, and one calibration screen (see Fig. 3). An encoder system delivers the actual slit position, and a second actuator serves for slit tilting with respect to the beam axis. Motor drivers and encoder electronics are made by Phytron GmbH. The distance between the slit mask and the YAG screen in screen-station 3 is 0.75 m. For image recording a 12-bit CCD camera Basler Scout with 659 x 494 pixels is used.

The corresponding size of one pixel on the screen is 25.3 µm. The control software is written in LabView and runs on a standard PC.

In the control software, the camera loop and the slit movement loop run in parallel, as shown in Fig. 4. The beamlet images are captured while the slit is continuously moving. In the $i$-th loop, the slit position $P_{ci}$ is recorded, typically at 10 Hz depending on the macro pulse trigger which indicates that beam is on. At the same time $T_{ci}$, a command is sent to the camera to capture one image. During the camera exposing, the recording time is $T_{ri}$. The camera exposure time is usually chosen as long as the macro pulse length to cover the full pulse and to reduce the background noise from the dark current. Taking the aforementioned into account, the exact position of the slit when the camera captures the image can be written as

$$P_{r_i} = \frac{P_{c_{i+1}} - P_{c_i}}{T_{c_{i+1}} - T_{c_i}} \cdot (T_{r_i} - T_{c_i}) + P_{c_i}. \tag{7}$$

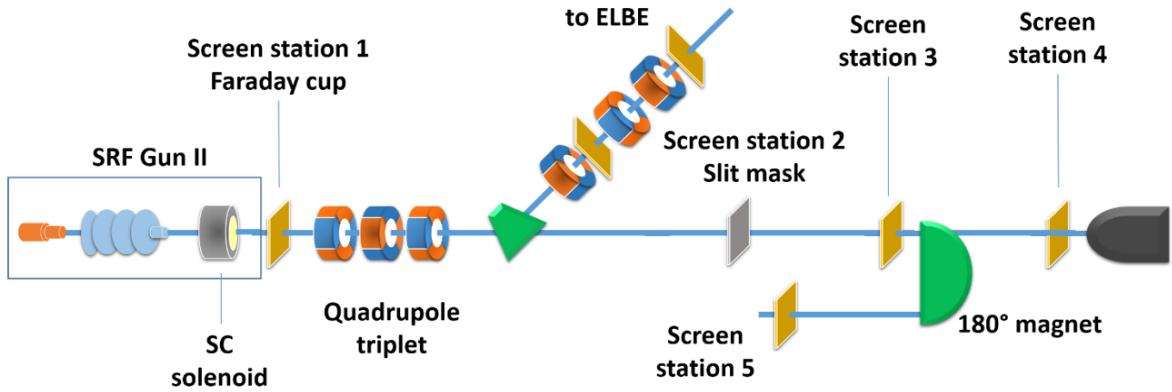

FIG. 2. The diagnostic beamline of ELBE SRF Gun II.

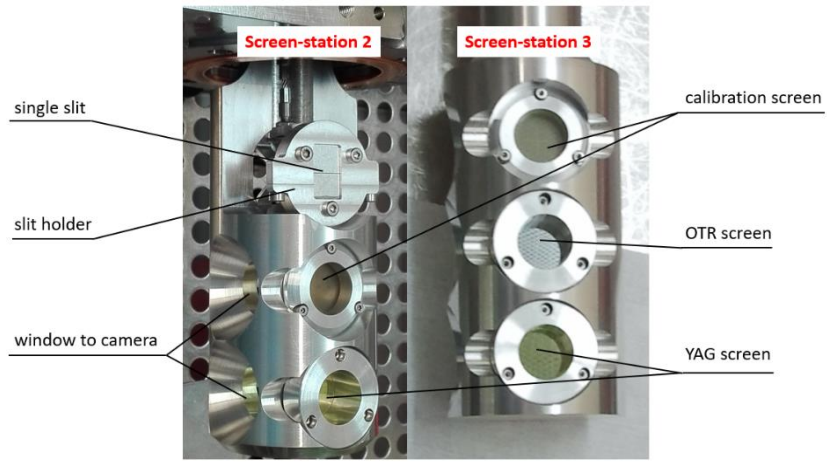

FIG. 3. Photographs of screen-station 2 and screen-station 3.

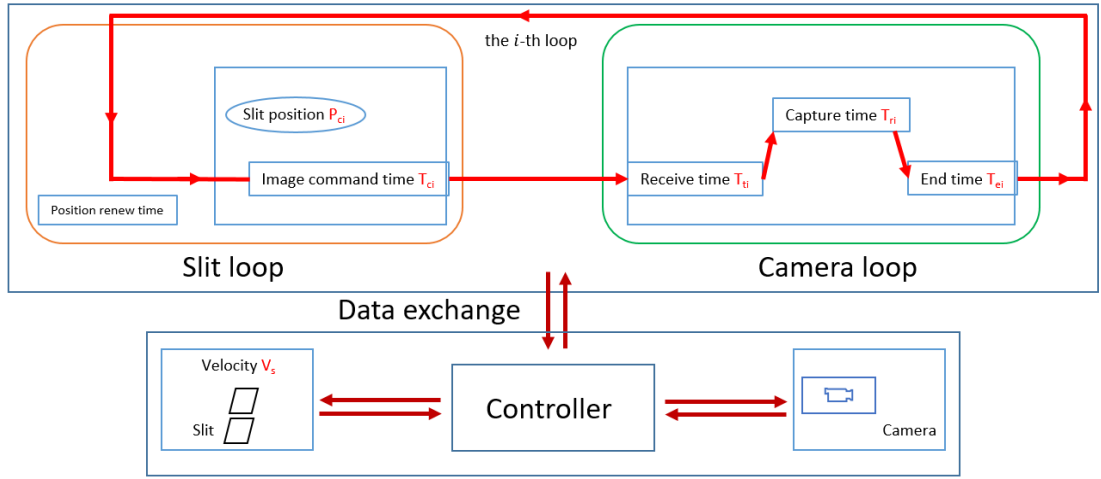

FIG. 4. Single-slit scan control system scheme.

## 3. Image processing based on Machine Learning

The first ML data processing step uses a CNN for image classification to eliminate datasets of needless images which contain no beamlet information. The input of the neural network is the horizontal projection of the beamlet images, which is a one-dimensional intensity array with 494 values. The neural network structure is shown in Fig. 5 and consists of two convolution layers, where one pooling layer, and a fully-connected layer with two nodes as the output layer. The output is a one-hot encoding to determine the processing image is useful for further treatment.

The second image processing step is mainly intended for noise elimination. The input data is the one-dimensional intensity array of the reduced number of beamlet image datasets selected in the first step. The output should be a nearly perfectly reconstructed beamlet intensity profile without noise and other artifacts. For that purpose, an encoder-decoder network is suitable [19]. Fig. 6 presents the general structure of such a network. The encoder-decoder consists of two parts, the encoder and the decoder. For the encoder part, the input is the data $x \in R^m$ and the output is the reconstructed data $y \in R^n$. One of the hidden layers stores the data features, written as the latent space $h_k$. The network has to learn the functions $f_e: R^m \to R^k$ and $f_d: R^k \to R^n$. In principle, the features of the input data will be learnt and stored into the latent space, then the decoder layers will rebuild the signal as the output from the latent space. The specific structure of the encoder-decoder network used in this paper is presented in Fig. 7.

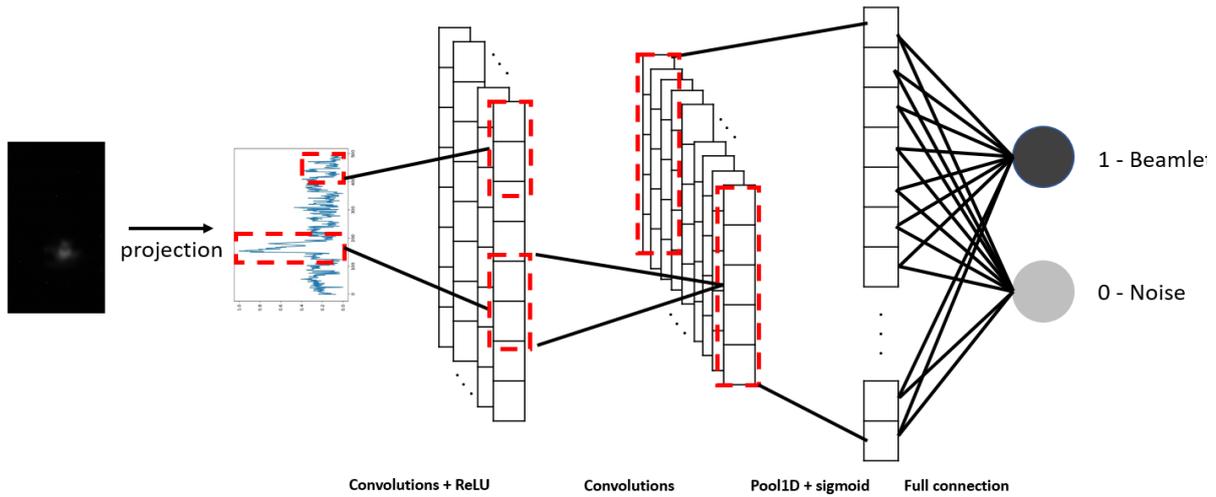

FIG. 5. Beamlet images classify network. The input data is a one-dimension signal with the beamlet intensity horizontal projection. The first convolution layer outputs 32 features with a leaky rectified-linear-unit (ReLU) function. The second convolution layer outputs 64 features. After this, a pooling layer and sigmoid function are used. The fully-connected layer is in the end and the log-softmax function scales the value from 0 to 1. The output value will be 1 when the log-softmax value between 0.5 and 1, and will be 0 otherwise.

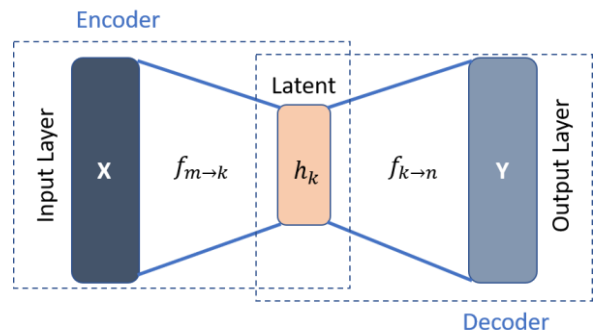

FIG. 6. Encoder-decoder network structure skeleton.

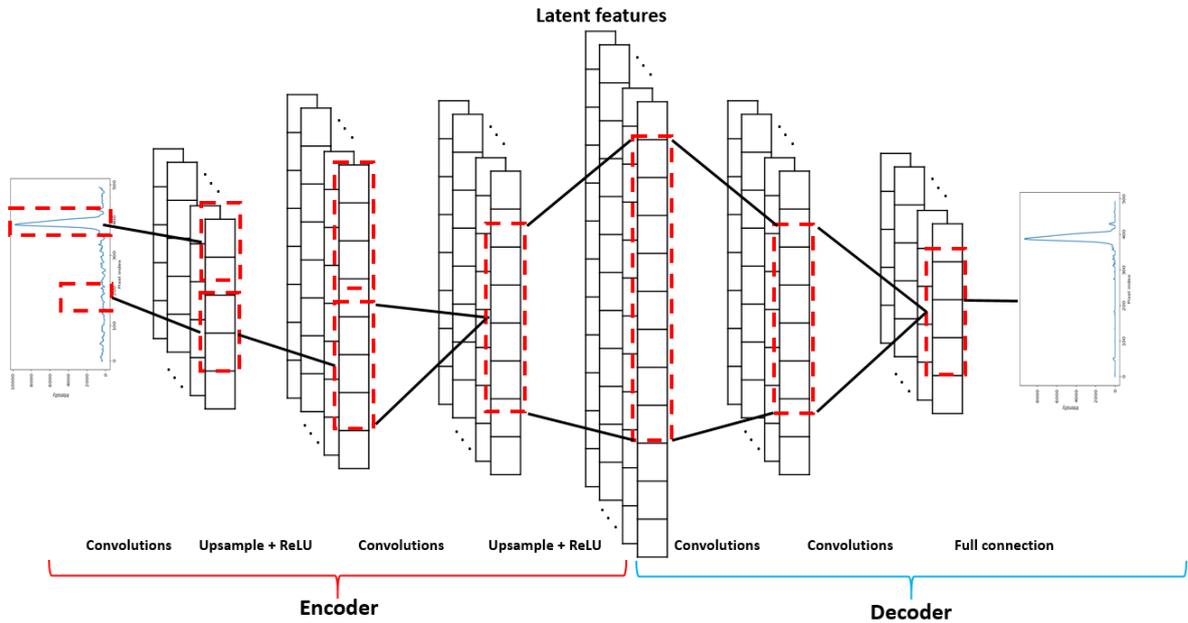

FIG. 7. The encoder-decoder network structures. The input data is the same as the classification network. The output features of the convolution layers in the encoder part are 16 and 32 with stride 1. The upsample layers are both nearest and scaling factors are 2. The latent features are 982. The output features of the convolution layers in the decoder part are 32 and 16 with stride 2. In the end is a full connected layer. The kernel sizes used are all 3. The total number of parameters is 2 005 887.

The ML networks have to be trained and tested. For the image classification network, 2500 projection datasets from beamlet images were used, which were taken from experimental slit-scan measurements. Thereby 2000 were utilized for training and 500 for testing. The training applies the stochastic gradient descent method with the cross-entropy [23] as a loss function. The final accuracy obtained from the test data is 98.8%. For the encoder-decoder network, the training data came from 1502 experimental beamlet images processed manually and through traditional filters. Noise signal data was taken from 107 experimental images without beamlet signals. Based on these, a random combination of filtered beamlet and noise data was used to construct 167205 projection data records, of which, in turn 80 % were used for training and 20 % for testing. Fig. 8 illustrates three examples. The training procedure was performed in the Maxwell Computer Cluster at DESY, using NVIDIA Tesla P100 GPU. The training time was about one and a half hours with one thousand epochs, 3072 batch size and mean square error (MSE) loss function. The optimizer used was Adam in Pytorch [24].

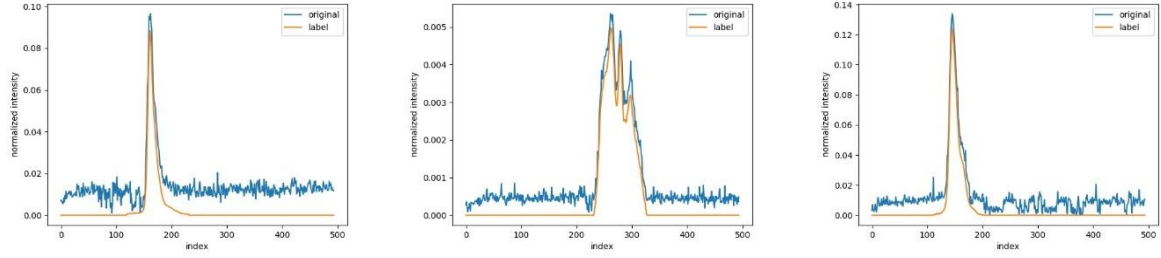

FIG. 8. The projection of beamlet images from simulation and the noise from the experiments.

Fig. 9 shows that the filter from machine learning (ML filter) has the potential to be more accurate than traditional filters, such as median and Butterworth filters. The true signal profile has two peaks with noise. The traditional filters can smooth the signal, but they are helpless to depress the noise to a low level. The Gaussian fitting method distort the signal if it is beyond the Gaussian distribution. The advantage of the ML filter is, that it doesn't need to adjust the parameters for different beamlet images once the neural network is well trained and evaluated.

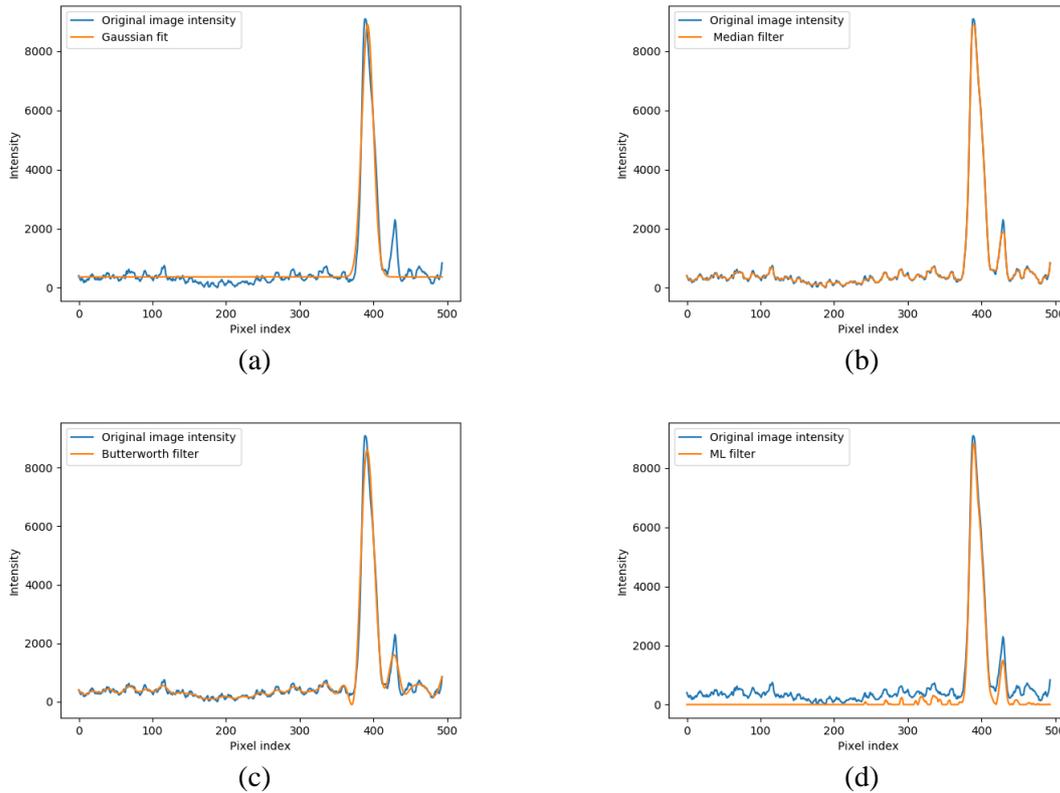

FIG. 9. One example of the comparison with different filters. (a) is Gaussian fitting; (b) is median filter with window length of seven; (c) is $5^{th}$ order Butterworth filter and critical frequency of 0.09; (d) is ML filter.

Although the ML filter can reduce noise efficiently, the noise from dark current will influence the beamlet position and the RMS size calculation. The slit-scan can be done twice to mitigate this influence in some extent. The first one is to measure the background from dark current, and the second scan includes beam and dark current. After subtracting the background, the further calculation of the beamlet center and RMS size are done twice. The first is by using pixel intensities of the full measurement range. Then in the second calculation, only intensities in a certain range, beamlet center plus and minus $f$ times RMS size from the first calculation, are considered. Usually, $f$ is chosen between 0.5 to 5 depending on different image conditions. Fig. 10 and Tab. 1 show an example calculated with different $f$ factors, which indicates that the beamlet RMS size is more sensitive to the value than the center position.

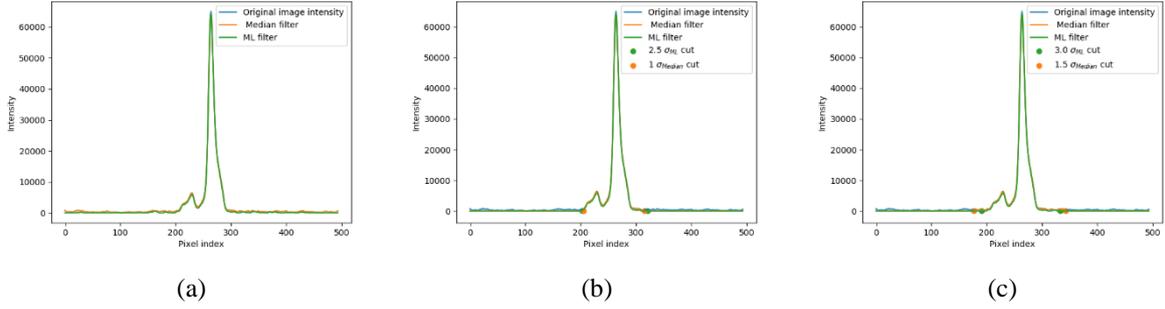

(a)                  (b)                  (c)

FIG. 10. One example with different cutting factors.

TAB. 1. Beamlets center, RMS size and intensity details of Fig. 17.

| | ML filter | | | Median filter | |
|---|---|---|---|---|---|
| Limit $f$ | Center (pixel) | RMS size (pixel) | Limit $f$ | Center (pixel) | RMS size (pixel) |
| 0 | 262.80 | 23.64 | 0 | 260.58 | 55.07 |
| 2.5 | 262.40 | 15.36 | 1 | 262.29 | 16.15 |
| 3.0 | 262.50 | 15.71 | 1.5 | 262.31 | 19.29 |

## 4. Emittance correction factor and error analysis

### 4.1. Emittance correction factor

The beam RMS size can also be measured directly using the YAG screen at station 2. Measurement results showed that the reconstructed beam sizes obtained from the beamlet measurements are mostly smaller than the beam size directly measured on the screen at the slit position. The reason is the finite signal-to-noise ratio which causes signal losses at the low-intensity edges of the beamlets. This effect has been identified for the first time at the PITZ photo injector [25], and a correction factor $f_c$ has been introduced:

$$f_c = \frac{\sigma_x}{\sqrt{\langle x^2 \rangle}}, \tag{8}$$

where $\sigma_x$ is the beam RMS size measured at slit position, and $\langle x^2 \rangle$ is the second central moment of the beam distribution as determined by the slits-can measurement. Then the corrected normalized emittance can be written as

$$\varepsilon_n = \frac{\sigma_x}{\sqrt{\langle x^2 \rangle}} \frac{p_z}{m_0 c} \sqrt{\langle x^2 \rangle \langle x'^2 \rangle - \langle xx' \rangle^2}. \qquad (9)$$

This conservative approach of the slit-scan analysis has been used in this paper and belongs to the standard slit-scan procedure at the ELBE SRF gun beamline. Fig. 10 shows the correction factor as a function of the bunch charge. For low bunch charges, the factor is in some cases smaller than one. The reason is the dark current or image noise, which is not subtracted from the signal completely, and by that enlarging the reconstructed beam RMS size. In most cases, the factor is between 1 and 1.1, especially for higher bunch charges.

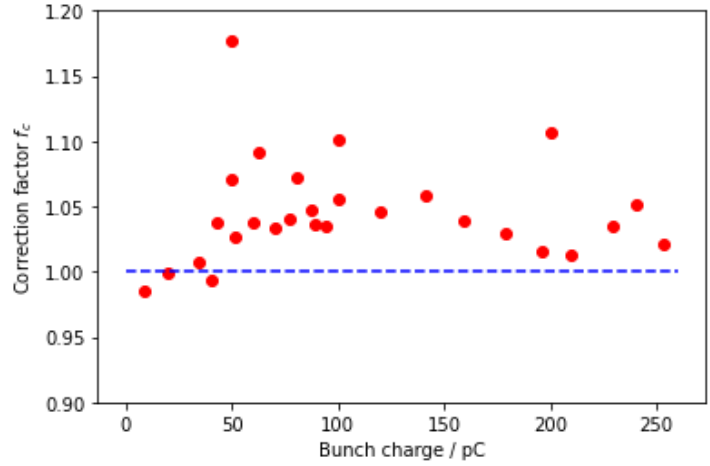

FIG. 10. Correction factor versus bunch charge. The blue dash line is equal to one and the red dots are the correction factors calculated from the slit-scan experiment at different bunch charges.

### 4.2. Error analysis

The accuracy of slit-scan emittance measurement is mainly determined by four sources: space charge influence in the beamlets ($e_1$), slit position recording error ($e_2$), uncertainties in the determination of the beamlet center position and RMS size ($e_3$), beam energy measurement uncertainly ($e_4$). The total relative error can be written as:

$$e_t = \sqrt{e_1^2 + e_2^2 + e_3^2 + e_4^2} \qquad (10)$$

Although the beam is divided into small slices which reduce bunch charge for each beamlet, the particle density of the beamlets is still the same, especially at the beginning of the drift space after the slit mask. If the beam density is large, the beamlets may suffer from space charge. Ref. [26] introduced a parameter based on the beam transverse envelope equation. Assuming that the beam distribution at the slit position is uniform and the slit width is $d$, the beamlet space charge dominance ratio is given by

$$R_b = \sqrt{\frac{2}{3\pi} \frac{I}{\gamma I_0} \left(\frac{d}{\varepsilon_n}\right)^2}. \qquad (11)$$

In this equation, $\varepsilon_n$ is the normalized beam emittance, $I$ is the beam peak current, and $\gamma$ is the Lorentz factor representing the beam energy and $I_0$ gives the characteristic current as $ec/r_e$. When $R_b \ll 1$, the beamlet is emittance dominated and the influence of space charge is negligible. Experimentally, it is sometimes impossible to fulfill this condition due to the technical limits of the screen visual scale and the camera sensitivity. To evaluate the magnitude of the error from space charge, we have done some series of slit-scan simulations considering the influence of the drift distance and space charge. In the simulation, the slit width is 100 µm and the step width is fixed as 100 µm, which can cover all beam without overlap. From the simulation results, the space charge and the drift distance cause errors in the emittance. Generally, when $R_b \leq 0.1$ and the drift distance is larger than 0.5 m, the error is less than 3% and will decrease with enlarging the drift distance, which is independent of space charge. For $R_b > 0.1$, the error from space charge increases with the drift distance enlarge, especially for large $R_b$, as shown in Fig. 11 (c).

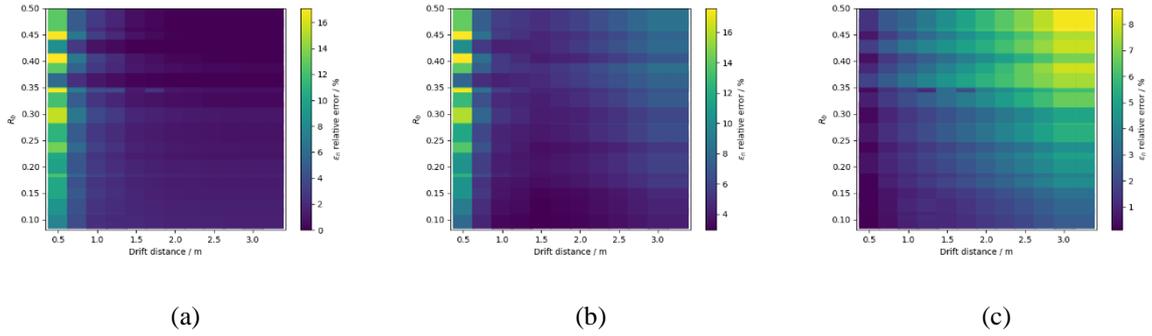

(a)          (b)          (c)

FIG. 11. Slit-scan simulations: (a) without space charge, (b) with space charge, (c) pure space charge, from (b) subtracts (a).

However, $R_b$ is related to beam parameters, such as bunch charge, beam spot size and bunch length. In the simulations, we have changed the laser power, laser spot size on the cathode, gun phase and solenoid current that all determine the aforementioned beam parameters. Fig. 12 shows how $R_b$ evolves with bunch charge for laser spot radii on the cathode of 1.25 mm and 1.875 mm. $R_b$ at smaller laser spot radius on the cathode is larger than the bigger one for bunch charge less than 75 pC. For high bunch charge, larger than 100 pC, the smaller laser spot on the cathode has smaller $R_b$. When $R_b$ is less than 0.5, the error is less than 10% from Fig. 11 (a).

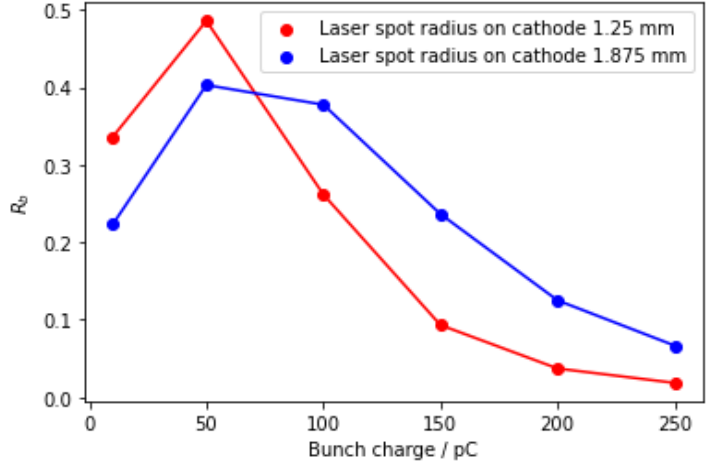

FIG. 12. $R_b$ versus bunch charge for two different laser spot radii on the cathode.

In our equipment, the slit recording uncertainty is less than 0.6% in average and the energy uncertainty is about 2%. These two factors are linear with the corresponding emittance relative error. Assume the averaged relative uncertainties of the beamlet intensity $\delta_n$, the center position $\delta_c$ and the square of RMS size $\delta_\sigma$, if they are independent for all beamlets and ignore the second and high order terms, the emittance relative error from these can be written as:

$$e_3 \approx \left|\delta_c + \delta_n + \frac{\delta_\sigma}{2}\right|. \tag{12}$$

The details about the derivation of this term are given in the Appendix A.

### 5. Experiment results

The experiments were carried out with beams of different bunch charges. The SRF gun gradient was set as 8.0 MV/m with a RF launch phase of 55°. The total energy of the beam was 4.45 MeV. As discussed in the former section, to subtract the dark current, every slit-scan was repeated twice where one was without beam and the other one was with beam with the aim to subtract the dark current. The total measurement time was one and a half minutes. Tab. 3 shows the detailed measurement errors at four bunch charges as examples. The correction factor $f_c$ was calculated by Eq. (8) and the influence of space charge, $e_1$, was taken from the slit-scan simulation. The slit position uncertainty from the motor system was 0.2 %. The beamlet image intensity, center and RMS size uncertainties were mainly from the system jitter and image noise. In reality, the beamlet from the beam center was to be an evaluation of average of the whole beamlets. Fig.11 shows examples of beamlet intensity, center and RMS size fluctuations at the screen position. The energy uncertainty was around 2 % in our facility due to magnet calibration and RF launch phase drift. The total error, $e_t$, is calculated by Eq. (8). The normalized emittance for different bunch charges is shown in Fig. 13.

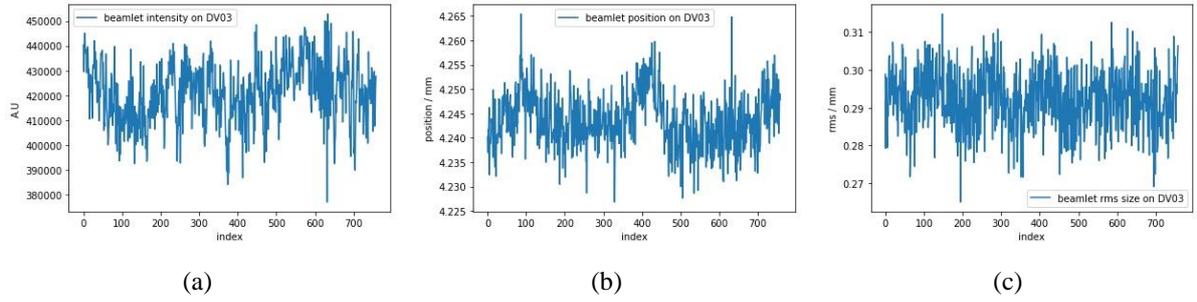

(a)                  (b)                  (c)

FIG. 13. Beamlet intensity, center and RMS size uncertainties at screen position. The bunch charge is 50 pC and the beamlet is from the center of the beam. The uncertainties of beamlet intensity, center and RMS size are about 2.9%, 0.9% and 10%.

TAB. 2. Examples of measured normalized emittances, beam size correction factors $f_c$ and errors.

| Bunch charge (pC) | $f_c$ | $e_1$ (%) | $e_2$ (%) | $e_3$ (%) | $e_4$ (%) | $e_t$ (%) | $\varepsilon_n$ (mm·mrad) |
|---|---|---|---|---|---|---|---|
| 50 | 1.07 | 7 | 0.2 | 8.8 | 2 | 18.0 | 1.90 |
| 100 | 1.06 | 8 | 0.2 | 5.0 | 2 | 15.2 | 3.65 |
| 159 | 1.04 | 5 | 0.2 | 4.9 | 2 | 12.1 | 4.04 |
| 200 | 1.11 | 3 | 0.2 | 2.5 | 2 | 7.7 | 4.29 |

As a comparison with normalized emittance from experiments, ASTRA simulations with different bunch charges were performed. The initial spatial beam distribution at the cathode is defined by two effects, the QE distribution of the cathode and the transversal intensity distribution of the laser. Two different emission conditions of the cathode have been appeared in the experiments, and are shown in Fig. 15. In the first row of Fig. 15, the upper plot (a) shows an inhomogeneous QE map of the cathode. With the laser of 3.75 mm diameter with uniform distribution shown in (b), it generates a beam distribution as in (c), which is applied in the simulation with the results showing the red line in Fig. 14. With a smaller laser spot size of 2.5 mm, the homogenous part of the QE distribution could be selected and it produces a beam with Gaussian distribution like in Fig. 15 (f). In Fig. 14, the green line presents the emittance results of this situation. Fig. 15 (d) shows a well-distributed QE mapping from the experiment. Together with an uniform laser shape as in Fig. 15 (e), the beam at the cathode has then a Gaussian distribution. The cyan line in Fig. 14 is the result from this simulation with a Gaussian distribution beam. In simulation and experiments the temporal distribution of the laser was Gaussian with 2.3 ps RMS pulse length. The results from the experiments and simulation agree well with each other.

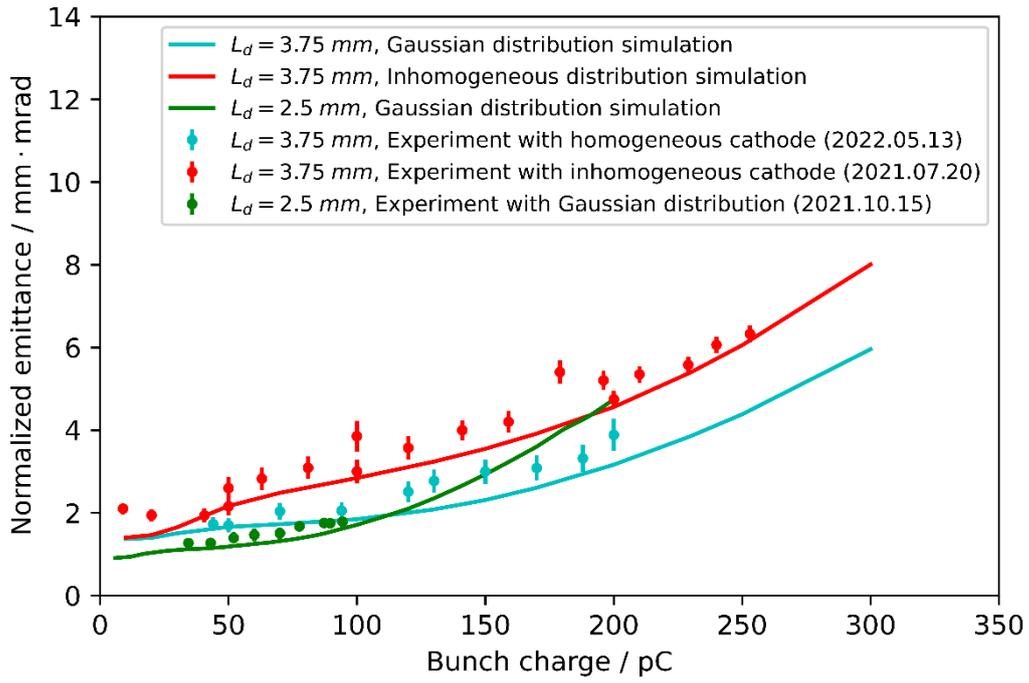

FIG. 14. Normalized emittance for varied bunch charges obtained from simulations and experiments with different spatial particle distributions at the cathode.

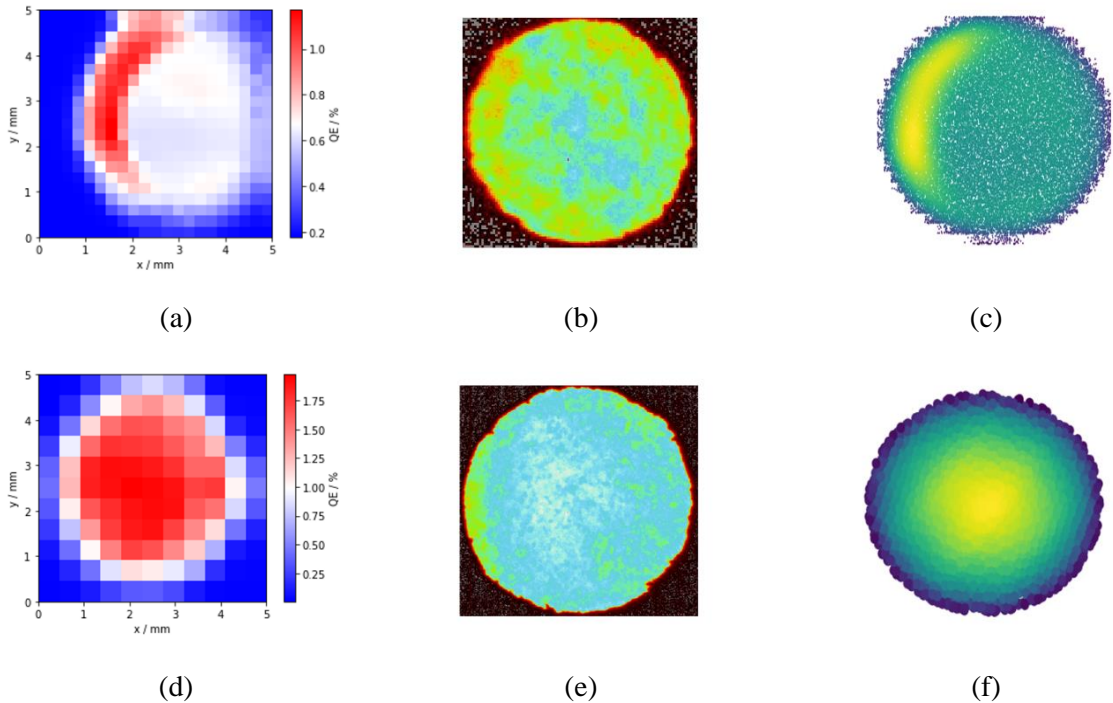

FIG. 15. (a) and (d) are the QE maps, (b) and (e) are the laser spot distributions on virtual cathode, (c) and (e) are original particles distributions in simulation. (c) is the inhomogeneous distribution and (f) is the gaussian distribution.

## 6. Conclusions

To improve the speed and the accuracy of the slit-scan emittance measurement at the ELBE SRF gun beamline, we constructed a new fast slit-scan system with a fast and continuous moving motor. Meanwhile, we have induced two machine learning algorithms, the CNN for classification and the encoder-decoder CNN for filtering noise to shorten the image processing time. The ML filter is potentially more effective than the traditional imaging processing methods for beamlet images. The system errors from the slit moving and the image processing were attentively analyzed. At the end, the experimental measurement results were compared with the ASTRA simulation. The results confirmed the preciseness of the new slit-scan system besides the significant improvement of measurement speed.

## Acknowledgement

We would like to thank the whole ELBE team for their help with the experiments. We also thank Dr. Houjun Qian from PITZ, DESY for his constructive discussions. This work was supported by China Scholarship Council and Chinese Academy of Engineering Physics. The machine learning part is supported by Helmholtz Information & Data Science Academy (HIDA) (No. 11952).

## References

[1] T. Rao and D. H. Dowell, An engineering guide to photoinjectors, arXiv:1403.7539 (2014).

[2] I. Grguraš, A. R. Maier, C. Behrens, et al. Ultrafast X-ray pulse characterization at free-electron lasers, Nature Photonics, 6(12): 852-857, 2012.

[3] S. Serkez, G. Geloni, S. Tomin, et al. Overview of options for generating high-brightness attosecond x-ray pulses at free-electron lasers and applications at the European XFEL, Journal of Optics, 20(2): 024005, 2018.

[4] C. Pellegrini, X-ray free-electron lasers: from dreams to reality, Physica Scripta 2016, 014004 (2017).

[5] L. Merminga, Energy recovery linacs, Synchrotron Light Sources and Free-Electron Lasers: Accelerator Physics, Instrumentation and Science Applications, 439-477, 2020.

[6] S. Wethersby, G. Brown, M. Centurion, T. Chase, R. Coffee, J. Corbett, J. Eichner, J. Frisch, A. Fry, M. Gühr, et al., Mega-electron-volt ultrafast electron diffraction at SLAC National Accelerator Laboratory, Review of Scientific Instruments 86, 073702 (2015).

[7] D. Kayran, Z. Altinbas, D. Bruno, M.R. Constanzo, A. Drees, A. V. Fedotov, W. Fischer, M. Gaowei, D. W. Gassner, X. Gu et al., High-brightness electron beams for

linac-based bunched beam electron cooling, Phys. Rev. Acc. and Beams, 23, 021003 (2020).

[8] C. Lejeune and J. Aubert, Emittance and Brightness: Definitions and Measurements in Applied Charged Particle Optics, Part A, edited by A. Septier (Academic Press, New York, 1980), p. 159 .

[9] J. D. Lawson, The Physics of Charged Particle Beams (Clarendon Press, Oxford, 1988), p. 156

[10] P. M. Lapostolle, Possible emittance increase through filamentation due to space charge in continuous beams, CERN Report ISR/DI-70-36 (1970); IEEE Trans. Nucl. Sci. NS-18, 1101 (1971).

[11] F. J. Sacherer, RMS envelope equations with space charge, IEEE Trans. Nucl. Sci. NS-18, 1105 (1971).

[12] T. Okugi, T. Hirose, H. Hayano, S. Kamada, K. Kubo, T. Naito, K. Oide, K. Takata, Seishi Takada, N. Terunuma, et al., Evaluation of extremely small horizontal emittance. Phys. Rev. ST Accel. Beams 2, 022801 (1999).

[13] H. Wiedemann, Particle Accelerator Physics (Springer Publishing, 2015) p. 224.

[14] K. M. Hock, M.G. Ibison, D. J. Holder, B.D. Muratori, A. Wolski, G. Kourkafas, B. J. A. Shepherd, Beam tomography research at Daresbury Laboratory. Nucl. Instrum. and Methods A 753, 38 (2014).

[15] M. Zhang, Emittance formula for slits and pepper-pot measurement, FERMILAB-TM-1988, Fermilab 1996.

[16] S. G. Anderson, J. B. Rosenzweig, G. P. LeSage, and J. K. Crane, Space-charge effects in high brightness electron beam emittance measurements, Phys. Rev. ST Accel. Beams 5, 014201 (2002).

[17] P. Michel, ELBE Center for high-power radiation sources, J. Large-Scale Res. Facil. 2, A39 (2016); https://www.hzdr.de/db/Cms?pNid=145.

[18] J. Teichert, A. Arnold, G. Ciovati, J.-C. Deinert, P. Evtushenko, M. Justus, et al., Successful user operation of a superconducting radio-frequency photoelectron gun with Mg cathodes, Phys. Rev. Accel. Beams 24, 033401 (2021).

[19] S. Indolia, A. K. Goswami, S. P. Mishra, and P. Asopa, Conceptual understanding of convolutional neural network-a deep learning approach. Procedia computer science, 132: 679-688 (2018).

[20] J. Zhu, Y. Chen, F. Brinker, W. Decking, S. Tomin, and H. Schlarb, High-fidelity prediction of megapixel longitudinal phase-space images of electron beams using encoder-decoder neural networks, Physical Review Applied, 16(2): 024005 (2021).

[21] D. P. Kingma, M. Welling, An introduction to variational autoencoders. arXiv preprint arXiv:1906.02691 (2019).

[22] K. Floettmann, A Space Charge Tracking Algorithm (ASTRA), http://www. desy.de/mpyflo (2007).

[23] K. P. Murphy, Machine learning: a probabilistic perspective (MIT press, 2012) p.54.

[24] Paszke A, Gross S, Massa F, Lerer A, Bradbury J, Chanan G, et al. PyTorch: An Imperative Style, High-Performance Deep Learning Library. In: Advances in Neural Information Processing Systems 32 [Internet]. Curran Associates, Inc.; 2019. p. 8024–35. Available from: http://papers.neurips.cc/paper/9015-pytorch-an-imperative-style-high-performance-deep-learning-library.pdf

[25] S. Rimjaem, G. Asova, J. Bähr, C. Boulware, H.J. Grabosch, M. Hänel, L. Hakobyan, Y. Ivanisenko, M. Khojoyan, G. Klemz, et al. Recent emittance measurement results for the upgraded PITZ facility[J]. Proceedings of FEL09, Liverpool, UK, 2009.

[26] S. G. Anderson, J. B. Rosenzweig, G. P. LeSage, J. K. Crane. Space-charge effects in high brightness electron beam emittance measurements. Phys. Rev. Accel. Beams, 5(1): 014201 (2002).

# Appendix A

Consider the averaged relative uncertainties of the beamlet intensity $\delta_n$, the center position $\delta_c$ and the square of RMS size $\delta_\sigma$. Assuming that they are independent for all beamlets, the $i$-th beamlet intensity, center and square RMS size can be written as

$$n_{i,r} = (1 \pm \delta_n)n_i, \tag{A.1}$$

$$\bar{x}_{sci,r} = (1 \pm \delta_c)\bar{x}_{sci}, \tag{A.2}$$

$$\sigma_{i,r}^2 = (1 \pm \delta_\sigma)\sigma_i^2, \tag{A.3}$$

where $\bar{x}_{sci,r}$ is the real center of the $i$-th beamlet on the screen. From these equations, $\langle x^2 \rangle_r$, $\langle x'^2 \rangle_r$ and $\langle xx' \rangle_r$ can be derived if we ignore higher order error terms:

$$\langle x^2 \rangle_r = \langle x^2 \rangle \pm \delta_n \langle x^2 \rangle, \tag{A.4}$$

$$\langle x'^2 \rangle_r = \langle x'^2 \rangle \pm \frac{2\delta_c}{L}\langle x'\bar{x}_{sc} \rangle \pm \delta_\sigma \langle \sigma'^2 \rangle \pm \delta_n \langle x'^2 \rangle, \tag{A.5}$$

$$\langle xx' \rangle_r = \langle xx' \rangle \pm \frac{\delta_c}{L}\langle x_s\bar{x}_{sc} \rangle \pm \delta_n \langle xx' \rangle. \tag{A.6}$$

From the equations above, it can be seen that coordinate systems of the slit and screen position cross each other, when the beamlet center has an error. Ignoring higher orders of $\delta_n$, $\delta_c$, $\delta_\sigma$, the calculated geometric emittance is

$$\varepsilon_c^2 \approx \varepsilon_0^2 \pm \frac{2\delta_c}{L}(\langle x^2 \rangle\langle x'\bar{x}_{sc} \rangle - \langle xx' \rangle\langle x_s\bar{x}_{sc} \rangle) \pm \delta_\sigma \langle \sigma'^2 \rangle\langle x^2 \rangle \pm 2\delta_n\varepsilon_0^2. \tag{A.7}$$

Introducing

$$\frac{\langle x^2 \rangle\langle x'\bar{x}_{sc} \rangle - \langle xx' \rangle\langle x_s\bar{x}_{sc} \rangle}{L} = k_c\varepsilon_0^2, \tag{A.8}$$

$$\langle \sigma'^2 \rangle\langle x^2 \rangle = k_\sigma\varepsilon_0^2, \tag{A.9}$$

then the emittance error from these uncertainties has the form

$$e_3 = \frac{|\varepsilon_c - \varepsilon_0|}{\varepsilon_0} \approx \left|\sqrt{1 \pm 2\delta_c k_c \pm 2\delta_n \pm \delta_\sigma k_\sigma} - 1\right| \approx \left|\delta_c k_c + \delta_n + \frac{\delta_\sigma k_\sigma}{2}\right|, \tag{A.10}$$

Note that $k_c$ and $k_\sigma$ depend on the distribution of the beam in phase space and on the relationship of the coordinate systems at the slit and screen position. Practically it is difficult to calculate them because they include the geometric emittance true values. Here we estimate them to be the same magnitude as the true emittance and be one. By this simplification, the error from beamlet intensity, center position as well as RMS size uncertainties is written as $\left|\delta_n + \delta_c + \frac{\delta_\sigma}{2}\right|$.